\documentclass[%
groupedaddress,
floatfix,
nofootinbib,
amsmath,amssymb,
prx
]{revtex4-2}

\usepackage{algorithm}
\usepackage{algpseudocode}
\usepackage{graphicx}
\usepackage[braket,qm]{qcircuit}
\usepackage[justification=justified]{subcaption}

\newcommand{\ketbra}[2]{\left\vert#1\right\rangle\left\langle#2\right\vert}
\newcommand{\inbraket}[3]{\left\langle#1\middle\vert#2\middle\vert#3\right\rangle}
\newcommand{\modulus}[1]{\left\vert#1\right\vert}
\newcommand{\tr}[1]{\text{tr}\left[#1\right]}
\newcommand{\floor}[1]{\lfloor #1\rfloor}
\newcommand{\set}[1]{\left\lbrace#1\right\rbrace}

\begin{document}

\title{Note on simple and consistent gateset characterization including calibration and decoherence errors}

\author{Jeffrey M. Epstein}
\affiliation{Atom Computing, Berkeley, CA}

\date{\today}

\begin{abstract}
Building high-fidelity quantum computers requires efficient methods for the characterization of gate errors that provide actionable information that may be fed back into engineering efforts. Extraction of realistic error models is also critical to accurate simulation and design of quantum circuits, including those used in quantum error correction. We provide a method for determining the parameters corresponding to decoherence in a model gateset. We demonstrate that this method is robust to SPAM and pulse area errors, and describe a simple and intuitive method for interpreting the quality and precision of the resulting error model.
\end{abstract}
\maketitle


\section{Introduction}

Achieving computational advantage with quantum processors will require the ability to perform gate operations with very high fidelity. A crucial component of the engineering process that enables progress towards this goal is a suite of robust and efficient characterization tools. Broadly speaking, characterization methods come in two varieties. Techniques such as randomized benchmarking (RB) \cite{Magesan2012} provide estimates of average gate fidelities, but do not result in explicit error models (beyond the simplest single-parameter models like pure depolarizing noise, which in many systems are not expected to be representative of real noise). Such methods are useful for tracking progress and comparing hardware, but do not provide directly actionable information. Other methods, such as robust phase estimation (RPE) \cite{Kimmel2015} and gateset tomography (GST) \cite{Nielsen2021}, provide estimates of errors in a particular error model, which may be directly related to physical sources of error in the particular realization of quantum hardware being analyzed. Such approaches allow both direct feedback into the engineering effort to improve error rates, as well as explicit prescriptions for simulation of arbitrary circuits, which may inform algorithm design.

In this note, we address the issue of characterizing gate level decoherence\footnote{This may include both ``intrinsic" decoherence, as measured by T2 times, and ``effective" decoherence resulting from classical noise in gates.}, taking the perspective of the second type of method. That is, our goal will be to estimate parameters in a well-defined, parameterized error model. As an example, we present a method for estimating two error parameters corresponding to decoherence in the $X$ and $Z$ bases during the implementation of a noisy $X90$ gate. We will also show that this method is robust against errors in state preparation and measurement (SPAM) as well as to errors in the driving angle. 

More general approaches to characterization of stochastic Pauli channels (of which single-qubit decoherence channels are a specific example) have been discussed previously \cite{Flammia2020, Flammia2021}, but have focused on the situation in which only an average error channel over a set of gates is considered, or in which only a single application of a gate is performed, so that the errors being examined are not amplified. To obtain a method that works well for characterizing errors associated to a specific gate, we combine the following intuitions: (1) repeated applications of a particular error channel allow a simple exponential fit to yield SPAM-robust estimates of error parameters; and (2) drive errors may be canceled with a phase flip in the middle of a long sequence, as in the rotary spin echo\cite{Solomon1959}.

As a demonstration of the use of this protocol, we describe a simple but effective procedure for quantifying the precision of the resulting estimates and the quality of the resulting error model. The statistical methods discussed for precision and model fit analysis are not independently novel, but here we show their utility applied to the characterization of quantum processors.

\section{Stochastic $X$ and $Z$ Error estimation}
An example universal gateset comprises active $X_{90}$ rotations, ``virtual" arbitrary angle Z-rotations \cite{mckay2017efficient} which are assumed to be error-free, and $CZ$ gates. Focusing first on 1Q gates, we would like a protocol to extract the parameters $p_x$ and $p_z$ corresponding to the strengths of decoherence in the $X$ and $Z$ bases during application of an $X_{90}$ gate. These parameters enter through the action of a channel
\begin{align}
\Lambda_{p_x,p_z}:\rho\mapsto \left(1-\frac{p_x}{2}-\frac{p_z}{2}\right)\rho + \frac{p_x}{2}X\rho X+ \frac{p_z}{2}Z\rho Z.\label{eq:decoherence_channel}
\end{align}
The output of such a protocol will be a model of the $X_{90}$ gate as a quantum channel $\tilde{\mathcal{X}}_{90}$ equivalent to a perfect unitary implementation $\mathcal{X}_{90}$ of the desired rotation followed by this decoherence channel:
\begin{align}
\tilde{\mathcal{X}}_{90}=\Lambda_{p_x,p_z}\circ\mathcal{X}_{90}\label{eq:noisy_gate}.
\end{align}
The choice of this convention for the error channel is somewhat arbitrary. For example, we could also choose to model the noisy gates as ideal gates followed first by a stochastic $Z$ error and then by a stochastic $X$ error, or vice versa. However, because to first order in the error rates $p_i$ the relation $\Lambda_{\mathbf{p}}\circ\Lambda_{\mathbf{q}}=\Lambda_{\mathbf{p}+\mathbf{q}}$ holds ($\mathbf{p}$ is shorthand for the pair $p_x, p_z$), this distinction is unimportant for small error rates. For the same reason, unless there are physical processes expected to produce stochastic $Y$ errors, we may reasonably model $p_y=0$. The protocol presented here shows that $X$ and $Z$ errors may be disentangled. In general, separate rates for each Pauli channel may not be learnable \cite{Chen2023}.

For the purposes of characterization, a key property of these channels is their simple representation in terms of permutations and scaling of Pauli matrices. In particular, the composite channel $\tilde{\mathcal{X}}_{90}^2$ is diagonal in the Pauli basis, with eigenvalues that are functions of $p_x$ and $p_z$. A reasonable strategy is thus to examine the decay of the Pauli operators under repeated application of the noisy channel. It is also desirable to isolate the decoherent part of the noise channel from potential coherent errors. A simple way to do so is to add an echo (a $Z180$ gate) in the middle of the sequence of repeated $X_{90}$ gates. Repeating this procedure for different circuit depths $m$, using Pauli $X$ and $Z$ operators as inputs and looking at their decays as a function of $m$ should allow us to fit an exponential whose decay parameter depends on the error parameters $p_x$ and $p_z$. Of course, the Pauli operators are not density matrices, so the linearity of quantum mechanics must be exploited to simulate the action of the circuits directly on these operators by taking linear combinations of measurement results. The resulting estimation procedure is as follows:

\begin{algorithm}[H]
\caption{Decoherence Detection}
\begin{algorithmic}
\Require a list $ms$ of even integers defining circuit depths
\Require an integer $K$ defining the number of shots per circuit
\Ensure estimates $\hat{p}_z$ and $\hat{p}_x$ of the decoherence parameters of a noisy $X90$ gate modeled by Eqs. \ref{eq:decoherence_channel} and \ref{eq:noisy_gate}
\\
\For{$P$ either of the Pauli operators $X$ or $Z$}
    \For{$m$ in $ms$}

        \For{$s=\pm 1$ }
            \State prepare the $s$ eigenstate of $P$
            \State apply the circuit $X_{90}^m-Z_{180}-X_{90}^m-Z_{180}$
            \State measure in the $P$ basis
            \State repeat $K$ times
            \State estimate probability $\text{Pr}(P, s, m)$ of outcome $s$
        \EndFor

        \State Define $S_P(m)=\sum_{s=-1}^1\text{Pr}(P, s, m)-1$
        
    \EndFor

    \State Fit $S_P(m)$ to the form $A_P\,\lambda_P^m+b_P$
  
\EndFor
\State Compute estimates $\hat{p}_z=1-\sqrt{\lambda_X}$ and  $\hat{p}_x=1-\sqrt{\frac{\lambda_Z}{1-\hat{p}_z}}$
\end{algorithmic}
\end{algorithm}
Under the assumption that $Z_{180}$ gates may be implemented perfectly (as is the case for platforms in which virtual $Z$ gates may be compiled into the phases of the following $X$ gates) and $X_{90}$ gates have errors of the type described above, the circuits used in this protocol are described by quantum channels
\begin{align}
\mathcal{N}_m&=\mathcal{Z}_{180}\circ\tilde{\mathcal{X}}_{90}^m\circ\mathcal{Z}_{180}\circ\tilde{\mathcal{X}}_{90}^m
\end{align}
and direct computation shows that these have the action
\begin{align}
\mathcal{N}_m(X)&=(1-p_z)^{2m}X\\
\mathcal{N}_m(Y)&=(1-p_x)^{m}(1-p_x-p_z)^mY\\
\mathcal{N}_m(Z)&=(1-p_x)^{m}(1-p_x-p_z)^mZ
\end{align}
on the Pauli matrices\footnote{These equations are essentially the nontrivial elements of the diagonal Pauli transfer matrix corresponding to the channel $\mathcal{N}_m$.}. Therefore in order to extract the parameters $p_x$ and $p_z$, we need to estimate the quantities
\begin{align}
S_P(m)=\tr{P\,\mathcal{N}_m(P)}/2
\end{align}
for $P=X$, $Z$ (normalized so that $P^2=I$). These values will obey the functional form
\begin{align}
S_P(m)=A_P\,\lambda_P^m+b_P,
\end{align}
where $\lambda_P$ is just the eigenvalue of the Pauli operator $P$ under the action of $\mathcal{N}_m$. By estimating this quantity for several values of the circuit depth $m$, an exponential decay may be fit, from which the error parameters can be extracted. Using the eigendecomposition of $P$ it may be shown that $S_P(m)$ is exactly the linear combination of probabilities used in the algorithm, demonstrating the correctness of the analysis\footnote{We have dropped second-order terms in $p_x$ and $p_z$ in order to obtain a simpler expression for the estimates.}.

\begin{figure*}[!ht]
    \centering
    \begin{subfigure}{0.35\textwidth}
    \includegraphics[width=\textwidth]{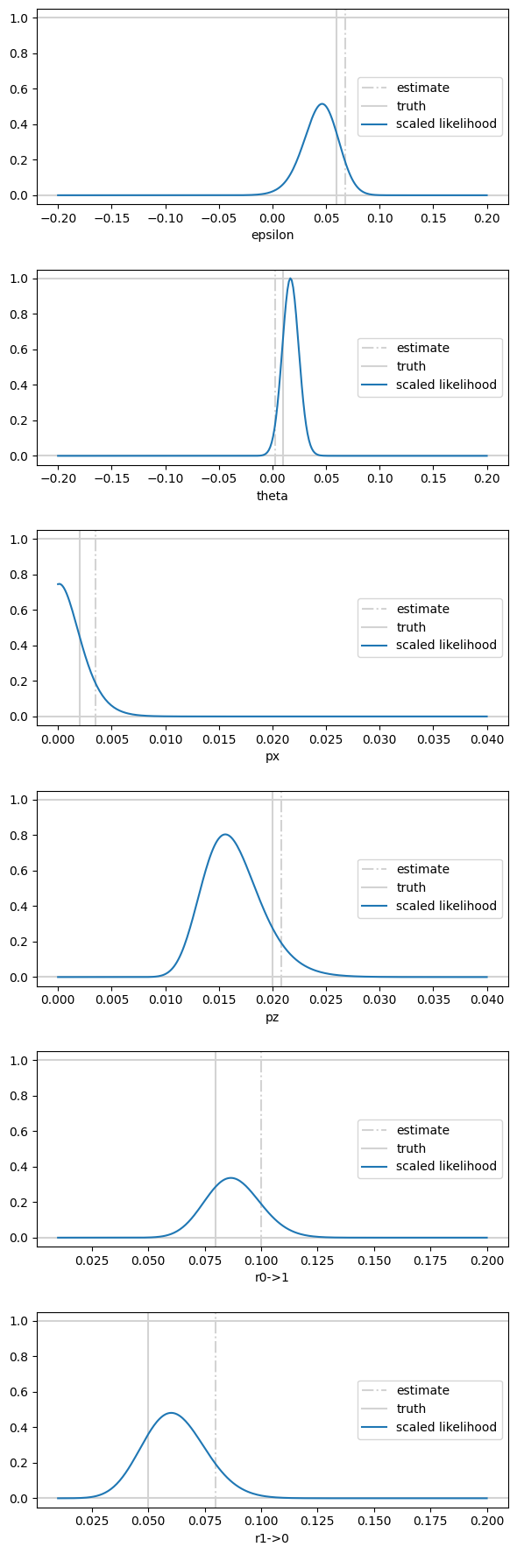}
    \end{subfigure}\hspace{30pt}
    \begin{subfigure}{0.35\textwidth}
    \includegraphics[width=\textwidth]{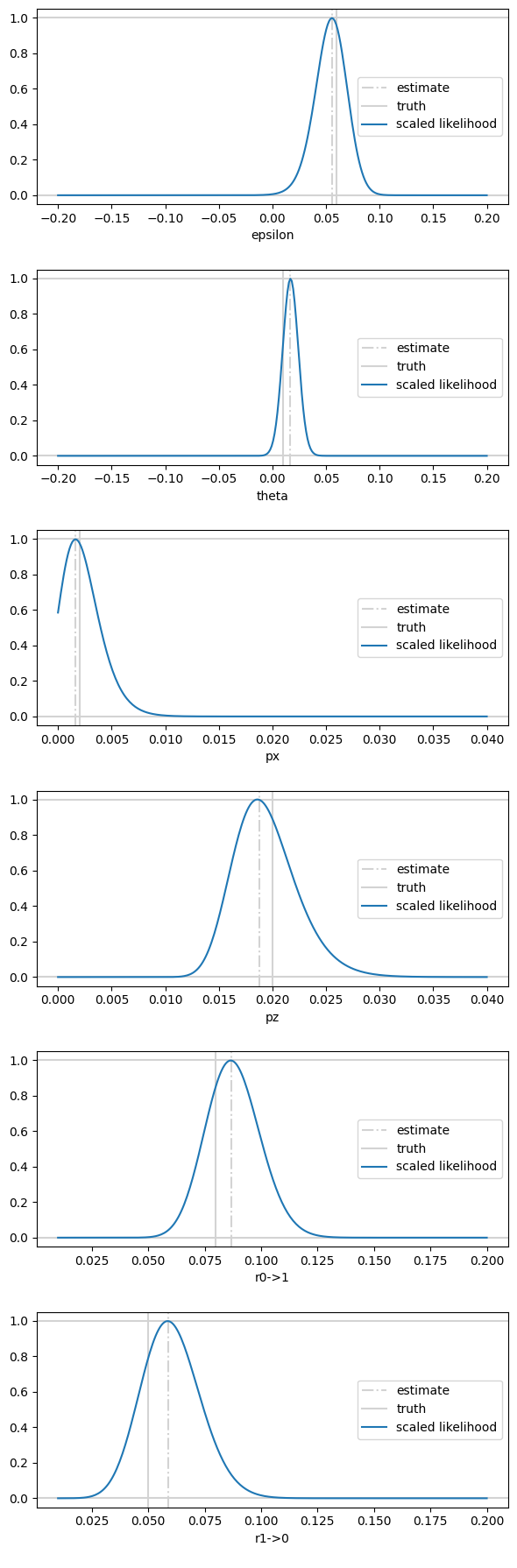}
    \end{subfigure}
    
    \caption{Precision analysis of a simulated set of characterization experiments and the resulting model estimate in the six-dimensional noise model space described in the text. On the left is the result of this analysis starting from the model obtained by collecting the independent estimates provided by the separate characterization protocols (RPE, decoherence detection, and readout error detection), and on the right the analysis is performed on the model resulting from likelihood maximization using the data from all of these circuits. Likelihoods are scaled separately on the left and on the right, but in fact the log-likelihood increases by about 5 between the estimate on the left and the one on the right. Note that the true values of the noise parameters lie within the likelihood peaks. The particular numerical values of the noise parameters are not representative of experimentally-observed values in any particular system, and are simply meant to demonstrate that the various types of errors (including different decoherence strengths in the $X$ and $Z$ bases) may be distinguished.}
    \label{fig:sim_plots}
\end{figure*}

It can be shown (Appendix \ref{app:SPAM}) that this algorithm is robust to SPAM errors in the sense that these errors only contribute to values of $A_P$ and $b_P$ differing from 1 and 0, respectively, but the form of $S_P(m)$ remains a single exponential decay with an unaltered decay rate. This robustness is exact, i.e. holds for any magnitude of SPAM errors, although for large errors of certain types $A_P$ may become very small, so that in practice fitting an exponential is difficult in the presence of statistical noise resulting from small sample sizes. Due to this robustness, the noisy $X90$ gate may be used in the eigenstate preparation and $X$ basis measurement steps of the protocol.

In addition, the algorithm is robust to errors in pulse area (i.e. over- or under-rotation about the $x$ axis) in the sense that for even values of $m$, these errors contribute only at second order to $S_P(m)$, while for odd values of $m$ they contribute at first order, but suppressed by the small parameter corresponding to measurement errors (Appendix \ref{app:pulse_area}). Thus overall, the method is robust to second order in errors affecting SPAM and pulse area. Unfortunately, the method is not robust against errors in rotation axis (detuning errors), which introduce oscillations in $S_P(m)$ so that in the presence of large detuning error, $S_P(m)$ must be fit to a decaying sinusoid rather than a single exponential.

\section{Approach to Gateset Model Estimation and Precision Analysis}
\label{section:stats}

Characterization and benchmarking approaches in quantum computing may be roughly organized along a continuum according to the amount of information they provide. On one end lie methods like randomized benchmarking (RB), which in it's most basic form provides a single number, an estimate of the average gate fidelity averaged over all Clifford gates\footnote{Even this quantity is not necessarily well-defined in the absence of an assumed form for state preparations and measurements, due to the unitary gauge invariance of quantum probability computations, see \cite{Proctor2017} for a thorough discussion of ambiguity in the interpretation of RB results.}. On the other are methods like full gateset tomography (GST), which provides full-rank estimates of the mathematical representations of all operational components of a quantum computation: density operators, corresponding to state preparation; process matrices, corresponding to gates; and POVM elements, corresponding to measurement outcomes\footnote{It is of course possible to characterize some kinds of errors that fall outside the Markovian, context-independent model. Methods for characterizing these types of noise fall beyond GST on the continuum described, or orthogonal to the axis.}. There are tradeoffs to be made in deciding where along this continuum to do characterization. RB, for instance, is simple and requires not too many measurements. The single number it reports, however, is not particularly informative, and gives no indication of how error rates may be reduced, nor does it allow prediction of the behavior of other circuits without further strong assumptions (purely depolarizing noise, for example). GST, on the other hand, is in principle maximally informative (within the setting of a Markovian, context-independent error model), but requires enormous amounts of data. Moreover, it is not clear that each individual matrix element of, e.g. a process matrix resulting from GST analysis provides actionable information.

With these considerations in mind, our approach is to work somewhere in between the endpoints of the Markovian characterization continuum, choosing a reduced model space\footnote{For a different, more mathematically-motivated type of reduced model space and accompanying model optimization framework based on the differential geometry of manifolds of gatesets, see \cite{Brieger2023}. While the compressive GST approach described in that reference is extremely elegant and will hopefully prove to be practically useful going forward, we have found it necessary to further restrict our gateset model spaces for efficient and interpretable characterization.} compared to GST that nonetheless provides actionable information for the further improvement of our gates. For single-qubit circuits, we choose the following model space:
\begin{enumerate}
\item State preparation is modeled by the ideal density matrix $\ketbra{0}{0}$. 
\item Rotations about the $z$ axis of the Bloch sphere, which are performed in software, are modeled by unitary channels corresponding to perfectly implemented rotations. Our X90 gates are modeled by rotations by an angle $(1+\epsilon)\pi/2$ about an axis lying at an angle $\theta$ above the equator, followed by a stochastic Pauli channel $\Lambda_{p_x,p_z}$.
\item Measurement is modeled by the POVM elements $M_0=(1-r_{01})\ketbra{0}{0}+r_{10}\ketbra{1}{1}$ and $M_1=(1-r_{10})\ketbra{1}{1}+r_{01}\ketbra{0}{0}$
\end{enumerate}
Thus our model space is parameterized by six error parameters. For the X90 gate, we have $\epsilon$ corresponding to under- or over-rotations, as would result from drive amplitude errors, $\theta$ corresponding to detuning errors, $p_x$ corresponding to dechoherence in the $X$ basis, as would result from shot-to-shot fluctuations in pulse area, and $p_z$ corresponding to decoherence in the $Z$ basis, as would result from phase noise in the control drives. Measurement errors are parameterized by $r_{01}$, corresponding to incorrectly reading out the state $\ket{0}$ as 1, and $r_{10}$, corresponding to the opposite error. Note that none of the approaches described here rely on the particular form of the error model.

With a model space fixed, we can perform experiments to obtain estimates of each of these parameters. RPE \cite{Kimmel2015} provides estimates of $\epsilon$ and $\theta$, the decoherence detection protocol described in the previous section provides estimates of $p_x$ and $p_z$, a simple experiment where $\ket{0}$ is prepared and measured provides an estimate of $r_{01}$, and $r_{10}$ is estimated by preparing $\ket{0}$, applying the X90 gate twice, and measuring. This set of experiments and analyses results in a point estimate of the error model as a point in the six-dimensional space described in the previous paragraph. Further refinement may be achieved by likelihood maximization in this model space using the results of all characterization circuits.

Once estimates of the error parameters are obtained, we have a complete gateset description of single-qubit circuits, which may be used to predict the outcomes of any single-qubit experiment we might choose to run, and from which the average gate fidelity of the noisy X90 gate may be computed. However, two natural questions remain. First, we would like to know how well this gateset explains the data taken in the course of its production. This will indicate whether the model ought to be extended to include errors of other types. Second, we would like to know how precise the experiments we performed are in terms of their ability to pinpoint exactly where in the model space our processor lies. If the experiments provide a quite precise estimate of the error parameters, the gateset model may be reasonably used to feed back into the engineering effort, performing whatever physical alterations are necessary to adjust each error parameter to zero. If on the other hand the estimates are quite imprecise, more data may need to be obtained before any action is taken to modify the hardware.

The first question may be addressed via a hypothesis testing type analysis. Briefly, we compute the statistic
\begin{align}
    \hat{\delta}&=\sum_i\modulus{\hat{p}_i-p_i},
\end{align}
where $\hat{p}_i$ is the empirical frequency of measuring zero after performing circuit $i$, and $p_i$ is the model probability of the same outcome given the gateset model obtained from the procedure described above. Intuitively, $\hat{\delta}$ should rarely be too large if the data is consistent with the model, and indeed this may be made precise, see Appendix \ref{app:hypothesis}.

The question of estimate precision may be addressed by looking at how the relative likelihood of models near the estimated maximum likelihood model falls off as each of the noise parameters is varied. A narrow peak in the likelihood function indicates a precise estimate. In particular, if the likelihood peak is well separated from zero, this should be taken as an indication that the correct course of action is to make hardware adjustments to bring that noise parameter closer to zero\footnote{In some cases, determining what the proper adjustments are may require significant work to understand how the experimental parameters directly controlled in the lab affect the model parameters. For single-qubit gates implemented in neutral atom platforms without complex pulse shaping, however, over- and under-rotations can be straightforwardly linked to pulse area errors, and incorrect rotation axis to detuning errors.}. Appendix \ref{app:likelihood} recapitulates the frequentist interpretation of this metric in terms of failure probabilities for distinguishing between two models given the observed data. Briefly, for each parameter we may define a meaningful interval characterized by the threshold likelihood
\begin{align}
\mathcal{L}_*=\frac{\mathcal{L}_\text{max}}{1/p_*-1},
\end{align}
where $\mathcal{L}_\text{max}$ is the likelihood, given the observed data, of the maximum likelihood estimate, and $p_*$ is a desired maximum failure rate for maximum likelihood discrimination between this estimate and another point in the model space. Models for which the likelihood exceeds this threshold should be included in the interval.

An example of this model fit and precision analysis approach is shown in Fig. \ref{fig:sim_plots}. The analysis is performed on data generated by simulating a gateset with error parameters $\epsilon=.06$, $\theta=.01$, $p_x=.002$, $p_z=.02$, $r_{01}=.08$, $r_{10}=.05$. These are not meant to be representative of physical error parameters, but are merely meant to illustrate the ability of these methods to distinguish noise due to different error parameters, in particular between decoherence in the $X$ and $Z$ bases. RPE up to depth 8 is used to estimate the coherent parameters, and the decoherence detection protocol described above is used with $m=20, 40, 60, 80, 100, 120$. For both of these, 30 shots per circuit are taken. For readout estimation, 300 shots are taken for each state preparation. The data from all of these circuits are then used in the likelihood maximization. For an experimentalist looking at the results, the takeaway should be that detuning and amplitude errors are likely both slightly positive, and that there is likely significant decoherence in the $Z$ basis, but insufficient evidence to conclude that there is decoherence in the $X$ basis. Readout errors are imprecisely quantified, but are well-demonstrated to be non-zero.

\section{Characterizing CZ Gates}
A similar simple model combining coherent and decoherent noise may be defined for CZ gates by modeling a noisy CZ channel as a diagonal unitary (i.e. with miscalibrated phases applied to each computational basis state), followed by decoherence in the Z basis:
\begin{align}
\tilde{\mathcal{CZ}}=\Lambda_\mathbf{p}\circ \mathcal{CZ},
\end{align}
where $\mathcal{CZ}$ is the channel corresponding to the unitary operator
\begin{align}
\begin{pmatrix}
1 & \\
& e^{i\alpha}\\
&& e^{i\alpha}\\
&&& e^{-(\pi+\beta)}
\end{pmatrix},\label{eq:unitaryCZerror}
\end{align}
which is a $CZ$ when the error parameters $\alpha$ and $\beta$ vanish, and $\Lambda_\mathbf{p}$ is a stochastic Pauli channel corresponding to the application of Pauli $P$ with probability $p_P$. This is a very restricted model of the noise that can affect two qubit gates, but is a reasonable first step in modeling entangling gates implemented via Rydberg interactions in neutral atoms, which do not necessarily drive transitions between computational basis states. The coherent error parameter $\alpha$ may be estimated by performing the circuits

\[
\Qcircuit @C=.6em @R=1.5em {
	& \lstick{\ket{0}}&\qw &\qw&\qw&\qw&\qw&\qw&\qw&\qw&\ctrl{1} &\ctrl{1}&\qw&\push{\rule{.3em}{0em}...\rule{.3em}{0em}}& &\ctrl{1} & \qw&\qw&\qw&\rstick{\ket{0}}\\
	& \lstick{\ket{0}}&\gate{H} &\qw&\qw&\qw&\qw&\qw&\qw&\qw&\ctrl{-1} &\ctrl{-1}&\qw&\push{\rule{.3em}{0em}...\rule{.3em}{0em}}& &\ctrl{-1} & \qw&\gate{H}&\qw&\rstick{\ket{0}}\\	
}
\]

\[
\Qcircuit @C=.6em @R=1.5em {
	& \lstick{\ket{0}}&\qw & \qw&\qw &\qw&\ctrl{1} &\ctrl{1}&\qw&\push{\rule{.3em}{0em}...\rule{.3em}{0em}}& &\ctrl{1} & \qw&\qw&\qw&\rstick{\ket{0}}\\
	& \lstick{\ket{0}}&\gate{H} &\qw&\gate{Z90}&\qw&\ctrl{-1} &\ctrl{-1}&\qw&\push{\rule{.3em}{0em}...\rule{.3em}{0em}}& &\ctrl{-1} & \qw&\gate{H}&\qw&\rstick{\ket{0}}\\	
}
\]
for $CZ$ depths $2^k$, $k=1,2,3,...$ and using the success probabilities to estimate the rotation angle of the state $\ket{01}$ with respect to the state $\ket{00}$ (the use of the two circuits related by a phase shift gives both quadratures of the rotation angle). Each value of $k$ gives the next bit of precision of this angle. The parameter $\beta$ may be estimated by preparing the first qubit in the state $\ket{1}$ to measure the relative phase applied between the states $\ket{10}$ and $\ket{11}$. As for single-qubit RPE, this estimation procedure is robust to SPAM noise.

If we model only $IZ$, $ZI$, and $ZZ$ stochastic Pauli errors\footnote{This is not the most general form of Z basis decoherence. Indeed, the decay rates of the coherences between each pair of basis states may be chosen independently as long as the rates form a valid correlation matrix, see \cite{Puchala2021}. This can be seen operationally by imagining implementing such a channel by flipping a set of correlated coins, one for each basis state, and applying a phase flip to each state whose coin lands on tails. This more general class of channels can be characterized using circuits analogous to the above, where superpositions of each pair of basis states are prepared, and the echo gate is a Pauli operator that permutes the two states.}, we can use the circuits
\[
\Qcircuit @C=.6em @R=1.5em {
	& \lstick{\ket{0}}&\gate{H} &\ctrl{1}&\ctrl{1} &\qw&\push{\rule{.3em}{0em}...\rule{.3em}{0em}}& &\ctrl{1}&\qw &\gate{X}&\qw&\ctrl{1}&\qw&\push{\rule{.3em}{0em}...\rule{.3em}{0em}}& &\ctrl{1} &\ctrl{1}&\gate{H}&\qw&\rstick{\ket{0}}\\	
	& \lstick{\ket{0}}&\gate{H} &\targ&\ctrl{-1}&\qw&\push{\rule{.3em}{0em}...\rule{.3em}{0em}}& &\ctrl{-1}&\qw &\gate{X}&\qw&\ctrl{-1} &\qw&\push{\rule{.3em}{0em}...\rule{.3em}{0em}}& &\ctrl{-1} & \targ&\gate{H}&\qw&\rstick{\ket{0}}\\
}
\]

\[
\Qcircuit @C=.6em @R=1.5em {
	& \lstick{\ket{0}}&\gate{H} &\qw&\ctrl{1} &\qw&\push{\rule{.3em}{0em}...\rule{.3em}{0em}}& &\ctrl{1}&\qw &\gate{X}&\qw&\ctrl{1} &\qw&\push{\rule{.3em}{0em}...\rule{.3em}{0em}}& &\ctrl{1} & \qw&\gate{H}&\qw&\rstick{\ket{0}}\\	
	& \lstick{\ket{0}}&\qw &\qw&\ctrl{-1}&\qw&\push{\rule{.3em}{0em}...\rule{.3em}{0em}}& &\ctrl{-1}&\qw &\qw&\qw&\ctrl{-1} &\qw&\push{\rule{.3em}{0em}...\rule{.3em}{0em}}& &\ctrl{-1} & \qw&\qw&\qw&\rstick{\ket{0}}\\
}
\]
and fit exponential decays to the success probabilities as the number of repeated CZ gates is increased. The decay rate for the first circuit gives $p_{IZ}+p_{ZI}$, and the second $p_{ZI}+p_{ZZ}$, and as in the single qubit case, the ``mirror" or ``echo" gates in the middle cancels out coherent errors of the form in Eq. \ref{eq:unitaryCZerror}.

These procedures are \textit{not} robust to errors in the $X$ gates, so we do not perform any detailed analysis here. However, note that the statistical analysis methods described in the previous section can be applied to the outcomes of these circuits to generate estimates of entangling gate error parameters by including these circuits and their outcomes with the X90 estimation protocols described above to obtain a gateset model including representations of state preparation, readout, X90, and CZ in a consistent way.

\section{Conclusion}
We have presented a method for characterization of decoherence errors in noisy implementations of an $X90$ gate, and demonstrated its use as part of a larger suite of tools that result in a Markovian error model for single-qubit circuits on a near-term quantum processor. These methods are simple and do not capture all possible types of error process, but in many cases capture the dominant errors faced by engineers working on early-stage quantum processors. Thus it is our hope that they provide a useful technique for rapidly obtaining actionable information about noise in experimental quantum computing platforms. In the future it will be interesting to develop modifications of the protocol that are also robust against detuning errors in X90 gates. \\
\\
\textbf{Acknowledgements.} The author thanks Ming Li for discussions on the statistical aspects of this approach and Eli Magidash for discussions on echo sequences. Thanks also to Josh Combes, Jonathan King, Kelly Ann Pawlak, David Rodriguez Perez, Miro Urbanek, and Evan Zalys-Geller for feedback on the text.

\bibliography{refs}

\clearpage
\appendix\begin{widetext}

\section{Robustness to SPAM Errors}\label{app:SPAM}
Consider a unital single-qubit quantum channel $\mathcal{E}$ diagonal in the Pauli basis:
\begin{align}
\mathcal{E}(P)=\lambda_P\,P.
\end{align}
Suppose that a noisy measurement in the $Z$ basis has POVM elements\footnote{For example, the readout errors modeled in Section \ref{section:stats} have $\delta M = r_{10}\ketbra{1}{1}-r_{01}\ketbra{0}{0}$.}
\begin{align}
M_0&=\ketbra{0}{0}+\delta M\\
M_1&=\ketbra{1}{1}-\delta M
\end{align}
and that noisy state preparations in the $Z$ basis are described by the density operators
\begin{align}
\rho_0&=\ketbra{0}{0}+\delta\rho_0\\
\rho_1&=\ketbra{1}{1}+\delta\rho_1,
\end{align}
where $\delta\rho_i$ is traceless and Hermitian. Then we have
\begin{align}
\text{Pr}_m\left(0\rightarrow 0\right)&=\tr{M_0\,\mathcal{E}(\ketbra{0}{0})}+\tr{M_0\,\mathcal{E}(\delta\rho_0)}\\
\text{Pr}_m\left(1\rightarrow 1\right)&=\tr{M_1\,\mathcal{E}(\ketbra{1}{1})}+\tr{M_1\,\mathcal{E}(\delta\rho_1)}
\end{align}
We have
\begin{align}
\tr{M_0\,\mathcal{E}(\ketbra{0}{0})}+\tr{M_1\,\mathcal{E}(\ketbra{1}{1})}&=\tr{M_0\,\mathcal{E}(\ketbra{0}{0})}+\tr{(1-M_0)\,\mathcal{E}(\ketbra{1}{1})}\\
&=\tr{M_0\,\mathcal{E}(Z)}+\tr{\mathcal{E}(\ketbra{1}{1})}\\
&=\tr{M_0\,\mathcal{E}(Z)}+\frac{1}{2}\tr{\mathcal{E}(1-Z)}\\
&=\tr{M_0\,\mathcal{E}(Z)}+1\\
&=\frac{1}{2}\tr{(1+Z+2\delta M)\,\mathcal{E}(Z)}+1\\
&=\left(1+\tr{\delta M\,Z}\right)\lambda_z+1\\
\end{align}
Using the tracelessness and Hermiticity of $\delta\rho_i$, we may write
\begin{align}
\delta\rho_i=r^i_x\,X+r^i_y\,Y+r^i_z\,Z
\end{align}
so that
\begin{align}
\tr{M_0\,\mathcal{E}(\delta\rho_0)}&=\inbraket{0}{\mathcal{E}(r^0_x\,X+r^0_y\,Y+r^0_z\,Z)}{0}+\tr{\delta M\,\mathcal{E}(\delta \rho_0)}\\
&=\inbraket{0}{r^0_x\,\lambda_x X+r^0_y\,\lambda_y Y+r^0_z\,\lambda_z Z}{0}+\tr{\delta M\,\mathcal{E}(\delta \rho_0)}\\
&=r_z^0\,\lambda_z+\tr{\delta M\,\mathcal{E}(\delta \rho_0)}.
\end{align}
and
\begin{align}
\tr{M_1\,\mathcal{E}(\delta\rho_1)}&=-r_z^1\,\lambda_z-\tr{\delta M\,\mathcal{E}(\delta \rho_1)}.
\end{align}
Combining these terms, we find
\begin{align}
\text{Pr}_m\left(0\rightarrow 0\right)+\text{Pr}_m\left(1\rightarrow 1\right)-1&=\left(1+\tr{\delta M\,Z}+r_z^0-r_z^1\right)\lambda_z+\tr{\delta M\,\mathcal{E}(\delta \rho_0-\delta\rho_1)}.
\end{align}
For vanishing SPAM errors, the linear combination of probabilities used in the algorithm discussed above thus yields exactly $\lambda_z$. Non-vanishing SPAM errors contribute at first order to an overall scaling, and at second order to an overall shift. Thus if a family of channels $\mathcal{E}_j$ obeys $\mathcal{E}_j(Z)=\lambda_j\,Z$, the algorithm produces estimates
\begin{align}
\hat{\lambda}_j=A\,\lambda_j+b
\end{align}
in the limit of large sample size, where $A$ and $b$ are constants depending only on the SPAM errors. The same analysis holds in the $X$ basis.

\section{Robustness to Pulse Area Errors}\label{app:pulse_area}
In the event of pulse area errors on the X90 gates in addition to stochastic Pauli errors, the X90 gate may be represented by the composite channel
\begin{align}
\mathcal{E}=\Lambda_{p_x,p_z}\circ\mathcal{X}_{90}^\epsilon,
\end{align}
where $\mathcal{X}_{90}^\theta$ is the channel corresponding to rotation by $\pi/2+\epsilon$ about the $x$ axis of the Bloch sphere. This channel has Pauli transfer matrix representation

\begin{align}
\mathcal{X}_{90}^\epsilon&=\begin{pmatrix}
1 & 0 & 0 & 0\\
0 & 1 & 0 & 0\\
0 & 0 &-\sin\epsilon & -\cos\epsilon\\
0 & 0 & \cos\epsilon & -\sin\epsilon
\end{pmatrix}
\end{align}
while the stochastic Pauli channel has the representation
\begin{align}
\Lambda_{p_x,p_z}&=\begin{pmatrix}
1 & 0 & 0 & 0\\
0 & \lambda_x & 0 & 0\\
0 & 0 & \lambda_y & 0\\
0 & 0 & 0 & \lambda_z
\end{pmatrix}.
\end{align}
The Z180 gate has the representation
\begin{align}
\mathcal{Z}_{180}&=\begin{pmatrix}
1 & 0 & 0 & 0\\
0 & -1 & 0 & 0\\
0 & 0 & -1 & 0\\
0 & 0 & 0 & 1
\end{pmatrix}.
\end{align}
Thus defining the channel
\begin{align}
\mathcal{N}_m&=\mathcal{Z}_{180}\circ\mathcal{E}^m\circ\mathcal{Z}_{180}\circ\mathcal{E}^m
\end{align}
it is easy to see that
\begin{align}
\mathcal{N}_m(X)&=\lambda_x^{2m}\,X.
\end{align}
Noting that $\mathcal{Z}_{180}^2$ squares to identity and commutes with the stochastic Pauli channel, we have
\begin{align}
\mathcal{N}_m&=\left(\Lambda_{p_x,p_z}\circ\mathcal{Z}_{180}\circ\mathcal{X}_{90}^\epsilon\circ\mathcal{Z}_{180}\right)^m\circ \left(\Lambda_{p_x,p_z}\circ\mathcal{X}_{90}^\epsilon\right)^m.
\end{align}
Restricting to the invariant subspace of all three channels spanned by $Y$ and $Z$, i.e. to the lower right block of the PTM representation, we have
\begin{align}
\left.\mathcal{N}_m\right\vert_{Y,Z}&=\left(\begin{pmatrix}
\lambda_y & 0\\
0 & \lambda_z
\end{pmatrix}\begin{pmatrix}
-\sin\epsilon & \cos\epsilon\\
-\cos\epsilon & -\sin\epsilon
\end{pmatrix}\right)^m\left(\begin{pmatrix}
\lambda_y & 0\\
0 & \lambda_z
\end{pmatrix}\begin{pmatrix}
-\sin\epsilon & -\cos\epsilon\\
\cos\epsilon & -\sin\epsilon
\end{pmatrix}\right)^m:=(DR^{-1})^m(DR)^m
\end{align}
Direct computation yields
\begin{align}
DR^{-1}DR&=\begin{pmatrix}
	\lambda_y\lambda_z& \lambda_y\epsilon(\lambda_y-\lambda_z)\\
	\lambda_z\epsilon(\lambda_y-\lambda_z) & \lambda_y\lambda_z
\end{pmatrix}+\mathcal{O}(\epsilon^2)\\
\\
DR^{-1}DR^{-1}DRDR&=(\lambda_y\lambda_z)^2\begin{pmatrix}
1& 0\\
0 & 1
\end{pmatrix}+\mathcal{O}(\epsilon^2)
\end{align}
so that by induction on $m$ we see that $\mathcal{N}_m$ is diagonal in the Pauli basis to first order for odd $m$ and to second order for even $m$. Thus for even $m$, we immediately see that pulse area errors do not contribute to errors in the extracted survival probabilities to first order, while for odd $m$, because the off-diagonal terms are only probed by errors in the measurement (i.e. a $Y$ component in the POVM elements of a $Z$ measurement), these errors are suppressed by another small parameter.

\section{Likelihood-based Precision Analysis}\label{app:likelihood}
For some sample space $\mathcal{X}$, consider two probability distributions $p_A$ and $p_B$. Consider the game in which one player chooses $A$ or $B$ uniformly at random, samples from the corresponding distribution, and passes the sample $x\in\mathcal{X}$ to the second player. This player is tasked with determining whether the sample comes from $p_A$ or $p_B$. If player 2 uses the maximum likelihood test, then the selected model is simply the one for which the probability assigned to $x$ is greater (or the result of a fair coin flip if the probabilities are equal). Thus, we can compute the probability, given observed outcome $x$, that the wrong determination is made. For a specific outcome $x$, let $p_>(x)$ denote the larger of $p_A(x)$ and $p_B(x)$, and define $p_<(x)$ similarly. Then
\begin{align}
\text{Pr}\left(\text{wrong}\vert x\right)=\frac{p_<(x)}{p_>(x)+p_<(x)}=\frac{1}{1+p_>(x)/p_<(x)}=\frac{1}{1+\mathcal{L}},
\end{align}
where $\mathcal{L}$ is the likelihood ratio. Thus, for example, if we pick a threshold of .05 for the error probability, we find that $\mathcal{L}>19$ is required. We can thus use likelihood ratios to examine the precision of our experiments. A small likelihood ratio compared to the model estimate obtained by likelihood maximization indicates that, given the observed data, our experiment has a high probability of correctly determining whether the data was drawn from the maximum likelihood distribution or some specific other distribution. It should be pointed out that while this analysis is computationally very simple, it does make contact with contentious issues in the foundations/philosophy of statistics, relying as it does on an application of the likelihood principle. Indeed, within the framework of likelihoodist statistics, this analysis might be considered to be backwards, deriving a non-fundamental frequentist interpretation of the likelihood ratio from the fundamental object, which is the likelihood function on the model space. From this point of view, the relative likelihood simply is the quantity that determines how much a certain experimental outcome favors one model over another.

\section{Model Deviation for Analytical Hypothesis Testing and Confidence Intervals}\label{app:hypothesis}
Given a particular gateset model $\mathcal{G}$ and a circuit $\mathcal{C}$, we have a probability distribution
\begin{align}
	P_{\mathcal{G},\mathcal{C}}:\mathcal{O}\rightarrow\mathbb{R}
\end{align}
over the set $\mathcal{O}$ of possible outcomes, i.e. over POVM elements in the gateset's measurement. Fixing a particular gateset $\mathcal{G}$ with a two-outcome measurement (let the outcomes be 0 and 1) and a set of circuits $\mathcal{C}_i$, $i=1,\ldots,N$, define
\begin{align}
	p_i=P_{\mathcal{G},\mathcal{C}_i}(1).
\end{align}
Consider performing a set of experiments where the distribution corresponding to $\mathcal{C}_i$ is sampled $n_i$ times. This results in a set of independent binomial random variables $\hat{X}_i=0, 1, \ldots,n_i$. From these, we can compute estimators
\begin{align}
	\hat{p}_i=\frac{\hat{X}_i}{n_i}
\end{align}
of the probabilities $p_i$. These estimators are again independent random variables.\\
\\
Suppose that we would like to determine whether or not a given set of data corresponding to $n_i$ runs of each circuit $\mathcal{C}_i$ was generated by the distributions $P_{\mathcal{G},\mathcal{C}_i}$. In general of course, this is impossible, since almost all gatesets are consistent with all outcomes. But intuitively it seems that if the empirical distributions $\hat{p}_i$ are quite far from the true distributions $p_i$ corresponding to a model, that model fails to be a good ``explanation" for that data. This leads to the following prescription for assessing model violation. Define the absolute deviations
\begin{align}
	\hat{\delta}_i=\modulus{\hat{p}_i-p_i}.
\end{align}
These are independent random variables with (generally nonzero) means and standard deviations
\begin{align}
	\mu_i&=\mathbb{E}\,[\hat{\delta}_i]\\
	\sigma_i&=\left(\mathbb{E}\,\left[\left(\delta_i-\mu_i\right)^2\right]\right)^{1/2}
\end{align}
Fortunately, these have analytical expressions \cite{stackexchange}:
\begin{align}
	\mu_i&=\frac{2}{n_i}(1-p_i)^{1-\floor{n_ip_i}}p_i^{\floor{n_ip_i}+1}(\floor{n_ip_i}+1)\binom{n_i}{\floor{n_ip_i}+1}\\
	\sigma_i&=\left(\frac{p_i(1-p_i)}{n_i}-\mu_i^2\right)^{1/2}
\end{align}
We can then consider the summed absolute deviation
\begin{align}
	\hat{\delta} = \sum_i\hat{\delta}_i
\end{align}
which has mean and standard deviation
\begin{align}
	\mu&=\sum_i\mu_i\\
	\sigma&=\left(\sum_i\sigma_i^2\right)^{1/2} 
\end{align}
Then for $k>0$, we have by Chebyshev's bound
\begin{align}
	\text{Pr}\left[\frac{\vert\hat{\delta} - \mu\vert}{\sigma}\geq k\right]\leq \frac{1}{k^2}.
\end{align}
We can now use this as a criterion for model violation, in the sense that, upon observing some experimental outcomes and computing $\hat{\delta}$, we can compute $\mu$ and $\sigma$ for a particular model, and evaluate the statistic
\begin{align}
	\hat{k}&=\frac{\vert\hat{\delta} - \mu\vert}{\sigma}.
\end{align}
We can then make the following statement: If data were drawn from the distributions defined by the gateset $\mathcal{G}$, the probability that we would see a value of $\hat{k}$ at least as large as observed is no greater than $1/\hat{k}^2$. Operationally, this means that if $1/\hat{k}^2$ is very small, we might want to look for a different model to ``explain" our data. An interesting feature of this approach is that, because the mean of the absolute deviation is non-zero, the model violation defined in this way is actually greater for a model that exactly matches the empirical probabilities than for one that is slightly off. This is unimportant, however, as the way to use this model violation is not to decide between models, but to determine whether to consider a model within the realm of possibility. Thus this model violation analysis will be used only to determine whether or not a model should be rejected, which we can do by throwing out any model which has $1/\hat{k}^2<.05$, to use a standard threshold.\\
\\
Suppose we are promised that data is drawn from the distributions corresponding to some unknown gateset $\mathcal{G}$ from a known family $\mathcal{M}$ of gatesets. From the observed outcome and for some $\epsilon>0$, determine the (subset-valued) random variable
\begin{align}
	\hat{R}_\epsilon=\set{\mathcal{G}'\in \mathcal{M}:1/\hat{k}^2>\epsilon}
\end{align}
By the analysis above, with probability $1-\epsilon$, $\mathcal{G}\in\hat{R}_\epsilon$, so taking for example $\epsilon=.05$ would make $\hat{R}_\epsilon$ a 95\% confidence region. In the setting of gateset characterization, we are certainly \textit{not} in this situation, as we do not have a fully parameterized space of all possible gatesets, so the best we can do is to say that a small $\hat{R}_\epsilon$ indicates that, were our system truly behaving as if it were governed by a gateset in the parameterized family under consideration, the measurement would be precise with high probability. Alternately, we can view $\hat{R}_\epsilon$ as the intersection of the actual $(1-\epsilon)$-confidence region in the very large (though finite-dimensional) space of all possible maps from circuits to distributions (still assuming i.i.d. RVs for each shot of each circuit). In practice, the resulting intervals are looser than those obtained by the relative likelihood approach discussed in Appendix \ref{app:likelihood}, so we use the latter.

\end{widetext}

\end{document}